\documentclass[conference]{IEEEtran}
\IEEEoverridecommandlockouts

\usepackage{cite}
\usepackage{amsmath,amssymb,amsfonts}
\usepackage{algorithmic}
\usepackage{graphicx}
\usepackage{textcomp}
\usepackage{xcolor}
\usepackage{mathrsfs}

\newcommand{\set}[1]{\mathcal #1} 
\usepackage{amsthm}
\theoremstyle{plain}
\newtheorem{theorem}{Theorem}
\newtheorem{lemma}{Lemma}

\newtheorem{corollary}{Corollary}

\newtheorem{definition}{Definition}
\newtheorem{remark}{Remark}
\DeclareMathOperator*{\argmin}{argmin}

\def\BibTeX{{\rm B\kern-.05em{\sc i\kern-.025em b}\kern-.08em
    T\kern-.1667em\lower.7ex\hbox{E}\kern-.125emX}}
\begin{document}

\title{On the Rényi Rate-Distortion-Perception Function and Functional Representations\\
\thanks{This work has been supported by the European Research Council (ERC) under the European Union’s Horizon 2020 research and innovation programme (Grant agreement No. 101003431).}
}

\author{ Jiahui Wei and Marios Kountouris\\
\small DaSCI, Department of Computer Science and AI, University of Granada, Spain \\
\small Email: \{jiahui.wei, mariosk\}@ugr.es
}

\maketitle

\begin{abstract}
We extend the Rate-Distortion-Perception (RDP) framework to the Rényi information-theoretic regime, utilizing Sibson’s $\alpha$-mutual information to characterize the fundamental limits under distortion and perception constraints. For scalar Gaussian sources, we derive closed-form expressions for the Rényi RDP function, showing that the perception constraint induces a feasible interval for the reproduction variance. Furthermore, we establish a Rényi-generalized version of the Strong Functional Representation Lemma. Our analysis reveals a phase transition in the complexity of optimal functional representations: for $0.5<\alpha < 1$, the coding cost is bounded by the $\alpha$-divergence of order $\alpha+1$, necessitating a codebook with heavy-tailed polynomial decay; conversely, for $\alpha > 1$, the representation collapses to one with finite support, offering new insights into the compression of shared randomness under generalized notions of mutual information.
\end{abstract}


\section{Introduction}
Information measures play a vital role in compression, transmission, and modern representation learning. While Shannon entropy, Kullback-Leibler (KL) divergence, and mutual information have traditionally dominated the field, they capture only the expected behavior of information density~\cite{polyanskiy2010, thomas2006elements}. 
In many operational scenarios, merely controlling the average case is insufficient. This necessitates generalizations to alternative information measures, among which Rényi entropy and the corresponding divergences \cite{renyi1961measures} have been widely established and studied in the information theory literature.
These measures provide an exponential view of coding cost (also known as Campbell's cost~\cite{CAMPBELL1965}) and allow for fine-grained control over tail distributions.
The Rényi framework has subsequently been extended to multiple domains and new problems, including robust hypothesis testing~\cite{csiszar_renyi_ht}, guessing~\cite{arikan_guessing, arikan_rd_guessing}, list coding~\cite{bunte_14_taskcoding}, among others~\cite{esposito2025sibson}.
This formulation provides a precise link to error exponents in channel coding~\cite{csiszar_renyi_ht} and cumulant generating functions in source coding~\cite{CAMPBELL1965, hill2025communicationcomplexityexactsampling}. 

In this work, we focus on Sibson's mutual information ($I_{\alpha}$)~\cite{Sibson1969} as a Rényi generalization. It is defined via a variational optimization over the marginal distribution $Q_Y$ and the Rényi divergence $D_\alpha(\cdot\|\cdot)$. Sibson’s mutual information satisfies essential properties such as the data processing inequality (DPI) and additivity\cite{esposito2025sibson}. It is worth noting that several alternative generalizations exist in the Rényi regime, including those due to Arimoto~\cite{arimoto_renyiMI}, Csiszár~\cite{csiszar_renyi_ht}, Lapidoth–Pfister~\cite{lapidoth_renyiMI}, and Hayashi~\cite{hayashi_renyiMI}. To date, there is no universally accepted definition of Rényi mutual information in the literature.

Classical rate-distortion theory~\cite{thomas2006elements} has recently been extended to the Rate-Distortion-Perception (RDP) framework~\cite{BlauMichaeliCVPR2018,Serra24}. This extension addresses the realism gap in generative modeling: strictly minimizing distortion measures (e.g., Mean Squared Error (MSE)) often leads to blurry and unrealistic outputs. The RDP framework introduces a perception constraint, $d_P(P_X, P_{Y}) \le \Delta$, which enforces that the generated marginal statistics remain aligned with the source distribution.

Nevertheless, current RDP analyses are predominantly Shannon-based. We argue that this can be limiting for modern applications involving distribution shifts or heavy-tailed data. By generalizing the RDP problem to the Rényi regime, we introduce a tunable parameter $\alpha$ that implicitly regularizes the learning objective~\cite{esposito2025sibson}. 
The regime $\alpha > 1$ penalizes large code lengths exponentially and enables control of the worst-case description length, whereas $\alpha < 1$ offers tunable robustness to outliers.

Beyond determining fundamental rate limits, it is equally vital to understand non-asymptotic performance and how such limits can be achieved. The Strong Functional Representation Lemma (SFRL)~\cite{li_gamal_sfrl, li_sfrl_tit} provides a constructive mechanism to simulate a target distribution $P_{Y|X}$ using explicit shared randomness and has been used in a wide range of source and channel coding problems~\cite{li_sfrl_tit, jwei25itw, hill2025communicationcomplexityexactsampling, Theis2021b}. We extend this structural perspective to the Rényi regime. By leveraging the Poisson functional representation~\cite{li_sfrl_tit} with respect to Sibson's optimal marginal~\cite{Sibson1969, verdu2015alpha}, we derive achievable bounds on the Rényi entropy of the source. This analysis uncovers a direct connection between the nature of the representation codebook and the order $\alpha$, revealing a phase transition between infinite and finite support that is invisible under standard Shannon analysis.

In this work, we rigorously formulate the Rényi Rate-Distortion-Perception problem and investigate the structural properties of its optimal solutions. Our main contributions are as follows. First, we define the Rényi RDP (R-RDP) function using Sibson’s $\alpha$-mutual information and interpret it as a convex optimization problem. For the scalar Gaussian case with MSE distortion and perception constraints, we derive a closed-form analytical expression for $R_{\alpha}(D, \Delta)$. Our proof leverages the sharp Young inequality~\cite{Barthe1997sharpyoung} to establish the converse, confirming that the affine Gaussian test channel remains optimal even in the Rényi setting. We further demonstrate that, for Gaussian sources, the perception constraint reduces to a feasible interval on the reproduction variance. We characterize the optimal solution as a universal optimizer, which can be viewed as a projection of the standard rate-distortion solution onto this feasible set, and we provide explicit expressions for the feasible intervals under several divergence constraints.
Finally, we prove a Rényi generalization of the SFRL and recover the classical result by taking the limit at $\alpha \to 1$. For the regime $\alpha \in (0.5, 1)$, we show that coding complexity is bounded by the $\alpha + 1$ divergence, implying an optimal representation with infinite support and heavy polynomial tails ($p_k \sim k^{-\frac{\alpha}{1-\alpha}}$). In contrast, for the regime $\alpha > 1$, we analyze the logarithmic moment and demonstrate that the representation collapses to finite support. 

The remainder of the paper is organized as follows. Section~\ref{sec_preliminaries} recalls the essential definitions in the Rényi regime and reviews RDP theory. Section~\ref{sec_rrdp} presents the R-RDP function and illustrates it with the Gaussian case, while Section~\ref{sec_r_sfrl} introduces the Rényi-generalized SFRL and presents simulation results.

\section{Preliminaries}
\label{sec_preliminaries}
\subsection{Notation}
Throughout this article, random variables and their realizations are denoted by capital and lower-case letters, respectively, \emph{e.g.}, $X$ and $x$.  
For a Polish metric space $\set{X}$, let $\left(\set{X}, \mathscr{B}\left(\set{X}\right)\right)$ be the
associated Borel measurable space induced by the metric. For two $\sigma$-finite measures $\nu$ and $\mu$ on $\set{X}$, we write $\nu \ll \mu$ to denote that $\nu$ is absolutely continuous with respect to $\mu$. The probability measure $P_X$ over $\left(\set{X} , \mathscr{B}\left(\set{X}\right)\right)$ denotes the distribution of $X$. 
For two probability measures $P$ and $Q$ with $P \ll Q$, $\frac{dP}{dQ}$ denotes the Radon-Nikodym derivative. When $P_X \ll \mu$, we denote by $p(x)$ the Radon-Nikodym derivative of $P_X$ with respect to $\mu$, which we refer to as the probability density function or probability mass function, as appropriate.
Finally, $\log(\cdot)$ denotes the base-2 logarithm throughout this article.

\subsection{Rényi information quantities}
We begin by recalling the fundamental information measures in the Rényi regime that generalize Shannon entropy, KL divergence, and mutual information.

\begin{definition}[Rényi Entropy~\cite{renyi1961measures}]
For a random variable $X\sim P_X$ and $\alpha \in (0, 1) \cup (1, \infty)$, the Rényi entropy is defined as
\begin{equation}
    H_{\alpha}(X) := \frac{1}{1-\alpha}\log \left( \int_\set{X} p(x)^\alpha d\mu(x)\right).
\end{equation}
\end{definition}

\begin{definition}[Rényi Divergence~\cite{renyi1961measures}]
For probability measures $P$ and $Q$ on a measurable space dominated by a $\sigma$-finite measure $\mu$, the Rényi divergence of order $\alpha$ is defined as
\begin{equation}
    D_{\alpha}(P\|Q) := \frac{1}{\alpha-1}\log \int_{\mathcal{X}} \left(\frac{dP}{d\mu}\right)^{\alpha} \left(\frac{dQ}{d\mu}\right)^{1-\alpha} d\mu.
\end{equation}
\end{definition}

Taking the limit as $\alpha \to 1$, Rényi entropy reduces to the standard Shannon entropy $H(X) = -\sum p(x) \log p(x)$ and $D_{\alpha}(P\|Q)$ converges to the KL divergence $D(P\|Q)$~\cite{erven_klrenyi}:
\begin{equation}
    D(P\|Q) := \int_{\mathcal{X}} \log \left( \frac{dP}{dQ} \right) dP.
\end{equation}

For the definition of Rényi mutual information, we adopt Sibson's formulation~\cite{Sibson1969, verdu2015alpha}, which relies on a variational minimization over the marginal distribution $Q_Y$.

\begin{definition}[Sibson's $\alpha$-Mutual Information~\cite{Sibson1969}]
For a joint distribution $P_{XY}$ with marginal $P_X$, the Sibson's $\alpha$-mutual information is defined as
\begin{equation}
    I_{\alpha}(X;Y) := \inf_{Q_{Y}} D_{\alpha}(P_{XY} \| P_X \times  Q_{Y}),
\end{equation}
where the infimum is taken over all probability measures $Q_Y$ on $\mathcal{Y}$ and is achieved by $Q_{Y_\alpha}$ given by
\begin{align}
\label{eq_sibson_qy}
    Q_{Y_\alpha}(y) = \frac{\left(\int_\set{X} p_X(x) p_{Y|X}^\alpha(y|x)d\mu(x)\right)^{1/\alpha}}{\int_\set{Y}\left(\int_\set{X} p_X(x) p_{Y|X}^\alpha(y|x)d\mu(x)\right)^{1/\alpha} d\mu(y)} 
\end{align}
\end{definition}

As $\alpha \to 1$, this quantity recovers the standard Shannon mutual information $I(X;Y)$.

\subsection{Rate-distortion and rate-distortion-perception theory}
We consider a source $X \sim P_X$ and a reproduction $Y$ generated by a transition kernel $P_{Y|X}$. The joint distribution is given by $P_{XY} = P_X P_{Y|X}$, and the reproduction marginal is $P_{Y}(y)=\int P_{Y|X}(y|x)dP_{X}(x)$.

A non negative measurable function $d: \mathcal{X} \times {\mathcal{Y}} \to [0, \infty)$ is a distortion measure with $d(x, y)=0$ when $y=x$. Moreover, a functional $d_P(P_X, P_{Y})$ is a perception measure if it is a divergence satisfying $d_P(P_X, Q) \geq 0$ with equality if and only if $Q = P_X$, and is convex in its second argument $Q$~\cite{BlauMichaeliCVPR2018}.

The rate-distortion perception function was proposed in~\cite{BlauMichaeliICML2019} and generalizes the conventional rate-distortion function. It is defined as follows
\begin{definition}[RDP function~\cite{BlauMichaeliICML2019}]
    The RDP function for $X$ with distortion and perception constraints $D, \Delta \geq 0$ is given by 
    \begin{equation}
        \begin{aligned}
        &R(D, P) := \inf_{P_{Y|X}}  I(X;Y) \\
        &\text{subject to:}\quad\mathbb{E}[d(X,Y)] \leq D, \quad d_P(P_X, P_{Y}) \leq \Delta .
        \end{aligned}
    \end{equation}
\end{definition}
By taking $\Delta \to \infty$, the RDP function reduces to the standard rate-distortion function $R(D)$.

\section{Rényi Rate-Distortion-Perception Function}
\label{sec_rrdp}
We now generalize the RDP framework to the Rényi regime by considering Sibson's mutual information $I_\alpha$ as the optimization objective.

\subsection{Problem formulation}
Let $X \sim P_X$ be a source random variable defined on $(\mathcal{X}, \mathscr{B}(\mathcal{X}))$. We consider a reproduction random variable $Y$ defined on the reproduction alphabet $\set{Y}$, generated via a stochastic kernel $P_{Y|X}$ from $\mathcal{X}$ to $\set{Y}$.
The joint distribution is given by $P_{XY} = P_XP_{Y|X}$, and the corresponding marginal distribution of $Y$ is denoted by $P_{Y}$.

We define the Rényi Rate-Distortion-Perception (R-RDP) function as the minimum Sibson's $\alpha$-mutual information required to satisfy both distortion and perception constraints. For a fixed order $\alpha \in (0, 1) \cup (1, \infty)$ and constraints $D, \Delta > 0$, the function is defined as
\begin{equation}
\label{eq_rrdp}
    \begin{aligned}
    &R_\alpha(D, \Delta) := \inf_{P_{Y|X}}  I_\alpha(X;Y) \\
    &\text{subject to:}\quad\mathbb{E}[d(X,Y)] \leq D, \quad d_P(P_X, P_{Y}) \leq \Delta .
    \end{aligned}
\end{equation}
where the infimum is taken over all stochastic (Markov) kernels $P_{Y|X}$.

\begin{remark}
The existence of a minimizing kernel and the and the optimal marginal $Q_{Y_\alpha}$ follows from the lower semicontinuity of Sibson's mutual information  $I_{\alpha}(X;Y)$ with respect to the weak topology on the space of probability measures~\cite{Csiszar1974Extremum}, together with the compactness of the feasible set under convex constraints. 
For continuous alphabets with unbounded distortion measures (e.g., MSE), compactness is ensured by the tightness of the feasible measures, which, by Prokhorov’s theorem~\cite{Prokhorov}, follows from the moment bounds induced by the distortion constraint $\mathbb{E}[d(X,Y)] \le D$ when the source $X$ has finite second moment.
\end{remark}

\subsection{Optimization properties and convexity}
The computation of the Rényi RDP function involves an optimization over the transition kernel $P_{Y|X}$. A fundamental question is whether this problem admits a unique solution and can be handled efficiently from a numerical standpoint. We analyze the convexity of the objective function $I_{\alpha}(X;Y)$ with respect to the kernel $P_{Y|X}$ for a fixed source distribution $P_X$.

For the case $0 < \alpha < 1$, Sibson's $\alpha$-mutual information $I_{\alpha}(X;Y)$ is known to be convex in $P_{Y|X}$~\cite{verdu2015alpha, esposito2025sibson}. Since the distortion constraint $\mathbb{E}[d(X,Y)] \leq D$ is linear in $P_{Y|X}$ and the perception constraint $d_P(P_X, P_{Y})$ is convex in $P_{Y}$ (and thus in $P_{Y|X}$), the RDP optimization problem is convex.
For $\alpha > 1$, the mutual information $I_{\alpha}$ is in general neither convex nor concave in the channel. However, we can employ a monotone convex surrogate, as shown in~\cite{verdu2015alpha}, given by:
\begin{equation}
J_{\alpha}(X;Y) := 2^{\frac{\alpha-1}{\alpha} I_{\alpha}(X;Y)}.
\end{equation}
The mapping $P_{Y|X} \mapsto J_{\alpha}(X;Y)$ is convex. Since the exponential function is strictly increasing, minimizing $I_{\alpha}(X;Y)$ is equivalent to minimizing $J_{\alpha}(X;Y)$. Hence, the Rényi RDP problem can be reformulated as a convex optimization problem in the surrogate domain. This guarantees that any local minimum is global and enables the use of alternating minimization algorithms, such as Blahut-Arimoto.

\subsection{Case study: Gaussian R-RDP function}
We now turn to the canonical case of a scalar Gaussian source $X \sim \mathcal{N}(0, \sigma_X^2)$ under squared-error distortion $d(x,y) = (x-y)^2$.

\subsubsection{The Gaussian solution}
We first state the result for the Rényi rate-distortion function (without perception constraint), which generalizes the classical rate–distortion function $R(D) = \frac{1}{2}\log(\sigma_X^2/D)$.

\begin{theorem}[Gaussian Rényi Rate-Distortion]
\label{thm_rd_gaussian}
    For a Gaussian source $X \sim \mathcal{N}(0, \sigma_X^2)$ and squared -error distortion $D \in (0, \sigma_X^2)$, the Rényi rate-distortion function is given by
    \begin{equation}
        R_{\alpha}(D) = \frac{1}{2}\log\left( 1 + \alpha \frac{\sigma_X^2 - D}{D} \right).
    \end{equation}
    The optimum is achieved by the standard affine Gaussian test channel $Y = cX + Z$, where $Z\sim \mathcal{N}(0, \sigma_Z^2)$ is Gaussian noise independent of  $X$.
\end{theorem}

\begin{proof}[Sketch of Proof]
\textit{Achievability:} We substitute the affine channel parameters $c = 1 - D/\sigma_X^2$ and $\sigma_Z^2 = D(1 - D/\sigma_X^2)$ into the definition of $I_{\alpha}$ and evaluate the resulting Gaussian integrals directly~\cite[Example~3.3]{esposito2025sibson}.
    
\textit{Converse:} The lower bound is established using the sharp Young inequality~\cite{Barthe1997sharpyoung}. By expressing $I_{\alpha}$ in terms of the $\alpha$-norm of the convolution of densities and applying the sharp inequality with optimal constants, it can be shown that the rate is lower bounded by a Gaussian reconstruction. The key step involves identifying that the extremizers of the sharp Young inequality are Gaussian functions. The details are provided in Appendix~\ref{apx:thm_rd_gaussian}.
\end{proof}

\subsubsection{Incorporating perception constraint}
When the perception constraint $d_P(P_X, P_{Y}) \leq \Delta$ is active, for Gaussian sources with an affine Gaussian test channel, the perception constraint reduces to a restriction on the reproduction variance, denoted by $\sigma_{Y}^2$. In particular, the constraint $d_P(\mathcal{N}(0, \sigma_X^2), \mathcal{N}(0, \sigma_{Y}^2)) \leq \Delta$ reduces to a one-dimensional condition on the variance ratio $t := \sigma_{Y}^2 / \sigma_X^2$, namely 
\begin{equation}
    \sigma_{Y}^2 \in [\underline{\sigma}_{Y}^2(\Delta), \overline{\sigma}_{Y}^2(\Delta)] \ \text{with} \ \underline{\sigma}^2_{Y}=\sigma_X^2t_{\min},\  \overline{\sigma}_{Y}^2=\sigma_X^2t_{\max}.
\end{equation}
For example, under the 2-Wasserstein metric, this interval is defined by $|\sigma_{Y} - \sigma_X| \leq \Delta$. We also provide explicit expressions for the feasible intervals under other divergence constraints like KL divergence, Rényi divergence in Appendix~\ref{apx:thm_rdp_gaussian}.

\begin{theorem}[Gaussian R-RDP function]
\label{thm_rdp_gaussian}
    Let $D\in (0, \sigma_X^2)$ and define $\sigma_{RD}^2 := \sigma_X^2 - D > 0$. The distortion-feasible interval (ensuring $\sigma_Z^2 \ge 0$) is given by $\mathcal{I}_D = \left[ (\sigma_X - \sqrt{D})^2, (\sigma_X + \sqrt{D})^2 \right]$, and the perception-feasible interval is given by $\mathcal{I}_P = \left[ \underline{\sigma}_Y^2(\Delta), \overline{\sigma}_Y^2(\Delta) \right]$.
    Whenever the feasible set $\mathcal{S} := \mathcal{I}_D \cap \mathcal{I}_P$ is nonempty, the optimal reproduction variance ${\sigma_Y^\star}^2$ is given by the Euclidean projection of $\sigma_{RD}^2$ onto $\mathcal{S}$. In this case, the Rényi RDP function is given by
    \begin{equation}
        R_{\alpha}(D, \Delta) = \frac{1}{2}\log\left( 1 + \alpha \cdot \text{SNR}({\sigma_{Y}^\star}^2) \right),
    \end{equation}
    where $\text{SNR}(\gamma) := \frac{c^2 \sigma_X^2}{\sigma_Z^2}$, with channel parameters $c := \frac{\sigma_X^2 + \gamma - D}{2\sigma_X^2}$ and $\sigma_Z^2 := \gamma - c^2\sigma_X^2$.
\end{theorem}

\begin{proof}
    The proof is provided in Appendix~\ref{apx:thm_rdp_gaussian}
\end{proof}
\begin{figure}
    \centering
    \includegraphics[width=0.6\linewidth]{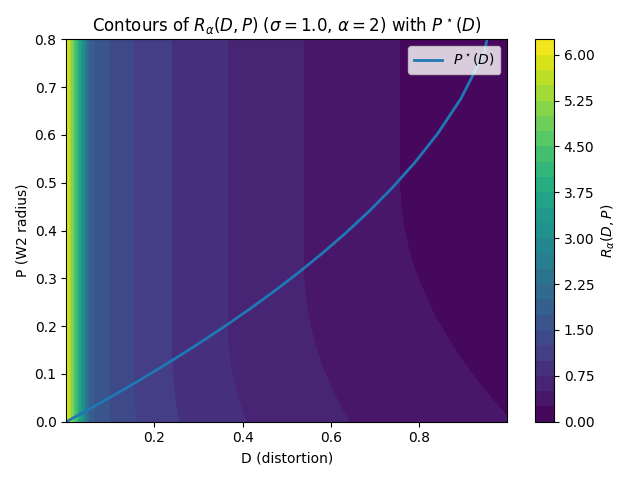}
    \caption{Contour plots of the Gaussian R-RDP function}
    \label{fig:fig_rrdp_gaussian}
\end{figure}
\begin{remark}
    Taking the 2-Wasserstein distance $W_2$~\cite{villani2008optimal} as a perception measure, we have $W_2(X, Y) = |\sigma_{Y} - \sigma_X | \le \Delta$, which yields $\underline{\sigma}_{}(\Delta)=\max\{0, \sigma_X - \Delta\}$ and $\overline{\sigma}_{Y}(\Delta)=\sigma_X + \Delta$ (as depicted in Fig.~\ref{fig:fig_rrdp_gaussian}). One can also show that taking the KL divergence as the perception measure reduces the problem to an optimization over the variance ratio parameter $t$, given $\Delta$. 
\end{remark}
The Rényi RDP function is minimized by choosing the variance $\sigma_{Y}^2$ closest to the standard RD-optimal variance $(\sigma_X^2 - D)$ that lies within the feasible interval; consequently, the optimum may be governed by either, or both, the distortion and perception constraints. The distortion-perception trade-off is illustrated in Fig.~\ref{fig:fig_rrdp_gaussian}, where we plot the contours of the Gaussian R-RDP function for $\sigma_X^2 = 1$ and $\alpha = 2$. The blue curve $P^\star(D)$ indicates the boundary at which the perception constraint becomes inactive for different Rényi communication rates.

\section{Rényi-Strong Functional Representation Bounds}
\label{sec_r_sfrl}
In this section, we extend the SFRL~\cite{li_gamal_sfrl} to the Rényi regime. We aim to construct a representation $Y$ of a given conditional distribution $P_{Y|X}$ using a discrete index $K$ and shared randomness $Z$, such that the coding cost, measured by the Rényi entropy of $K$, is upper bounded by Sibson's mutual information. We employ the Poisson point process construction introduced in~\cite{li_sfrl_tit}, with respect to the Sibson-optimal marginal distribution $Q_{Y_\alpha}$ given in~\eqref{eq_sibson_qy}.

\begin{theorem}[Rényi-SFRL]
\label{thm_renyi_sfrl}
Let $(X,Y) \sim P_{XY}$ be random variables on $\mathcal{X} \times \mathcal{Y}$. 
There exists shared randomness $Z$, independent of $X$, and a deterministic mapping $g: \mathcal{X} \times \mathcal{Z} \to \mathcal{Y}$ such that $Y = g(X, Z)$ almost surely.
Furthermore, the coding cost, measured by the Rényi entropy $H_\alpha(K)$ of the index $K$, satisfies the following bounds:

\begin{itemize}
    \item Case $\alpha \in (0.5 ,1)$: The coding cost is bounded via the $\alpha$-moment\footnote{For $\alpha \in (0, 0.5]$, the optimal distribution with infinite support is not normalizable under the $\alpha$-moment constraint alone, and therefore requires additional constraints.}:
    \begin{equation}
    \label{eq_renyi_1}
        H_\alpha(K) \leq \frac{1}{1 - \alpha}\log\left( \frac{\lambda}{\alpha}(2^{\alpha I_{\alpha+1}(X;Y)}+\alpha) + \frac{\mu}{\alpha} \right).
    \end{equation}
    \item  Case $\alpha > 1$: The coding cost is bounded via the logarithmic moment:
    \begin{equation}
    \label{eq_renyi_2}
    \begin{aligned}
        H_\alpha(K) \leq \frac{1}{1-\alpha} \log &\left(-\frac{\lambda}{\alpha} \left(I(X; Y) + D(P_Y||Q_Y)+ 1\right)\right. \\
        &- \left.\frac{\mu}{\alpha}\right).
    \end{aligned}
    \end{equation}
\end{itemize}
Here, $\lambda$ and $\mu$ are Lagrange multipliers satisfying the moment constraints derived in the proof.
\end{theorem}
Although expression~\eqref{eq_renyi_2} contains negative terms, the argument of the logarithm is strictly positive. Indeed, it is equivalent to the minimized power sum $\sum_{k} p_k^\alpha$, which is strictly positive for any valid probability distribution (see Eq.~\eqref{eq_pk_2}).

\subsection{Proof: Construction via Poisson Functional Representation}
We employ the Poisson functional representation using the minimizing distribution $Q_Y$. Let $Z = \{(\tilde{Y}_i, T_i)\}_{i \geq 1}$ be a Poisson point process on $\mathcal{Y} \times \mathbb{R}_{\geq 0}$ with intensity measure $Q_Y \times \nu$, where $\nu$ denotes the Lebesgue measure. Conditioned on $X = x$, we select the index $K$ as 
\begin{equation}
K := \argmin_{i} \frac{T_i}{\frac{dP_{Y|X}(Y_i|x)}{dQ_Y(Y_i)}}.
\end{equation}
By the displacement theorem applied to Poisson processes~\cite{kingman1992poisson}, setting $Y = \tilde{Y}_K$ ensures that $Y \sim P_{Y|X}(\cdot|x)$. Thus, $Y$ is a measurable function of $X$ and the shared randomness $Z$ (indexed by $K$). Note that we do not take $Q_Y = P_Y$; instead, we choose $Q_Y = Q_{Y_\alpha}$ in order to characterize $I_\alpha(X;Y)$.

\subsection*{Rényi entropy bounds}
For $\alpha\in (0, 1)$, the function $f(u) = u^\alpha$ is concave. We invoke the generalized Poisson matching lemma~\cite{li_sfrl_tit}. For any concave, non-decreasing function $g$, we have the stochastic dominance relation $\mathbb{E}[g(K-1)] \leq \mathbb{E}_P[g(\frac{dP}{dQ})]$. Applying this result with $g(u) = u^\alpha$, we obtain
\begin{equation}
    \mathbb{E}[K^\alpha \mid X=x] \leq 2^{\alpha D_{\alpha+1}(P_{Y|X}(\cdot|x)||Q_Y)} + \alpha.
\end{equation}
Taking expectation with respect to $P_X$~\cite{verdu2015alpha}, together with the choice $Q_Y = Q_{Y_{\alpha}}$, we use the identity $I_{\alpha + 1}(X; Y) = D_{\alpha+1}(P_{XY}||P_XQ_{Y_\alpha})$ to obtain the global moment constraint
\begin{equation}
    \mathbb{E}[K^\alpha] \leq 2^{\alpha I_{\alpha + 1}(X; Y)} + \alpha \triangleq C_\alpha.
\end{equation}
This moment bound is directly connected to the variational characterization of Sibson's $\alpha$-mutual information.

For $\alpha \in (1, \infty)$, the function $u^\alpha$ is convex, which precludes a direct upper bound via the Poisson matching lemma. Instead, we utilize the logarithmic moment $\mathbb{E}[\log K]$, as in the conventional Shannon case. 
Applying the Poisson matching lemma yields
\begin{equation}
    \mathbb{E}[\log K] \leq I(X;Y)+D(P_Y||Q_Y) + 1 \triangleq C_{log}.
\end{equation}

The final step consists of maximizing the Rényi entropy subject to the previous exponential or logarithmic moment constraints. Solving this optimization problem yields 
\begin{align}
    &p_k^{\alpha-1} = \frac{\lambda k^\alpha + \mu }{\alpha} \quad \text{for } \alpha\in (0.5, 1), \\
    \text{and}\ &\label{eq_pk_2}
    p_k^{\alpha-1} =\frac{-(\lambda \log k + \mu)}{\alpha} \quad \text{for } \alpha\in (1, \infty).
\end{align}
Here, $\lambda$ and $\mu$ are Lagrangian multipliers. 
Note that for $\alpha \in (0, 1)$, we have the asymptotic behavior $p_k \sim k^{\frac{\alpha}{\alpha-1}}$. In order for the distribution ${p_k}$ to be normalizable, we further require $\frac{\alpha}{1- \alpha} > 1$ (i.e., $\alpha > 0.5$), which guarantees that $\sum_k p_k$ converges to $1$.
The full derivation is provided in Appendix~\ref{apx_sfrl}.


The optimal distribution maximizing Rényi entropy under the $\alpha$-moment constraint has infinite support with a polynomially decaying tail ($p_k \sim k^{-\frac{\alpha}{1-\alpha}}$). This implies that representations in the regime $\alpha < 1$ are inherently heavy-tailed, requiring a conceptually infinite codebook to achieve the information-theoretic limit.
A critical question is whether this heavy-tailed behavior is merely an artifact of the Poisson point process construction or whether it reflects a more fundamental structural property of channel simulation under R\'enyi constraints.
We argue that, for $\alpha \in (0.5,1)$, this heavy-tailed behavior emerges from the variational structure of the R\'enyi-entropy optimization, rather than solely from the Poisson point process construction. More precisely, the infinite-support conclusion follows from maximizing $H_\alpha(K)$ subject to the derived $\alpha$-moment bound: because the function $u^\alpha$ is concave for $\alpha<1$, the entropy-maximizing distribution under this constraint is spread over an infinite support and exhibits polynomial decay, rather than collapsing to a finite alphabet.
Within the functional representation, the likelihood ratio
\[
L_i = \frac{P_{Y|X}(Y_i|x)}{Q_{Y_\alpha}(Y_i)}
\]
governs the selection of the index $K$. Under a stringent perception constraint $\Delta$, the generated marginal statistics must remain tightly aligned with the source distribution. When the source exhibits rare or atypical events, the likelihood ratio $L_i$ can fluctuate significantly. To simulate the target distribution $P_{Y|X}$ while respecting the perception budget, the optimal marginal $Q_{Y_\alpha}$ must assign sufficient mass to these rare events.
For $\alpha<1$, large index values are penalized only sub-linearly through the $\alpha$-moment criterion. Accordingly, the representation absorbs these likelihood fluctuations by developing a heavy polynomial tail, which suggests that an effectively infinite codebook is needed to attain the information-theoretic limit. As $\alpha \to 1$, this polynomial tail becomes progressively lighter; at $\alpha=1$, the classical Shannon case is recovered via the logarithmic-moment (Zipf-type) characterization.

In contrast, for $\alpha > 1$, the optimal distribution under the log-moment constraint has strictly finite support. In particular, the probability $p_k$ vanishes for all $k > K_{\max}$, where $K_{\max} \approx 2^{I(X;Y)+ D(P_Y||Q_Y)}$. This implies that the strong functional representation effectively compresses the source into a finite alphabet, whose size is governed by the mutual information (up to a divergence penalty).

\begin{remark}
A direct application of the Poisson matching lemma to the $\alpha$-moment is precluded by the convexity of $x^\alpha$ for $\alpha > 1$. Instead, we bound the logarithmic moment, corresponding to the Shannon capacity. Since $H_\alpha(K) \le H(K)$ for $\alpha > 1$, this approach provides a valid, albeit potentially loose, upper bound. 
\end{remark}

\begin{remark}
Taking the limit $\alpha \to 1$, we have $Q_{Y_\alpha} \to P_Y$, and the Rényi-SFRL reduces to the Shannon case in~\cite{li_sfrl_tit} by employing a Zipf distribution for the index $K$.
\end{remark}

\subsection{Link to Campbell's coding cost}
The moment bound derived for $\alpha \in (0, 1)$ admits a direct interpretation in terms of Campbell's coding cost. Recall that for a code with codeword lengths $\ell(k)$, the Campbell cost of order $t$ for binary codes is defined as $L(t) := \frac{1}{t} \log \mathbb{E}[2^{t \ell(K)}]$. A fundamental result due to Campbell states that for $t > 0$, this cost is sandwiched by the Rényi entropy of order $\alpha = \frac{1}{t+1}$, namely $H_\alpha(K) \leq L(t) < H_\alpha(K) + 1$. Combining this inequality with the bound on $H_\alpha(K)$ from Theorem~\ref{thm_renyi_sfrl}, we obtain a corresponding bound on the cost of coding the functional representation. 

\begin{corollary}[Campbell's cost bound]
For $0 < t < 1$ and $\alpha = \frac{1}{t + 1}$, the operational Campbell coding cost of the functional representation $K$ under the distortion-perception constraints is bounded as
\begin{equation}
\begin{aligned}
    L(t) & <  H_\alpha(K) + 1 \\
    &\leq \frac{1}{1 - \alpha}\log\left( \frac{\lambda}{\alpha}(2^{\alpha R_{\alpha+1}(D, \Delta)}+\alpha) + \frac{\mu}{\alpha} \right) + 1.
\end{aligned}
\end{equation}
\end{corollary}

\begin{remark}
The authors of~\cite{hill2025communicationcomplexityexactsampling} also employ the Poisson functional representation to analyze the Campbell cost of exact sampling for all $t > 0$. In contrast, our bound explicitly characterizes the cost of the RDP transition kernel in terms of Sibson's mutual information $I_{\alpha+1}(X;Y)$ (equivalently $R_{\alpha+1}(D, P)$), highlighting the complexity-distortion-perception trade-off. Further work is required to extend these results to $t\geq 1$ (i.e., $\alpha \le 0.5$).
\end{remark}

\subsection{Numerical simulation}
\label{subsec_simulation}
To validate the theoretical bounds derived in Theorem~\ref{thm_renyi_sfrl} and to build intuition for the functional representation in the Rényi regime, we perform numerical simulations on a scalar Gaussian source.
A key objective is to verify that the coding cost is upper bounded by Sibson's mutual information when the codebook is generated according to $Q_{Y_\alpha}$.
We consider a standard Gaussian source $X \sim \mathcal{N}(0, 1)$ with distortion constraint $D = 0.4$ and perception constraint $P = 0.1$. We evaluate two regimes, $\alpha = 0.6$ and $\alpha = 2$. For $\alpha = 0.6$, this yields $R_{\alpha+1}(D,P) \simeq 0.912$ bits.

The Poisson functional representation is approximated using a finite codebook of size $N = 10^5$.
For each source realization $x$, the index $K$ is selected by minimizing the ratio between the arrival times and the likelihood ratio, i.e., $K = \arg\min_i \frac{T_i}{L_i}$, where $L_i = P(Y_i|x)/Q_{Y_\alpha}(Y_i)$.

\begin{figure}[ht]
    \centering
    \includegraphics[width=0.47\linewidth]{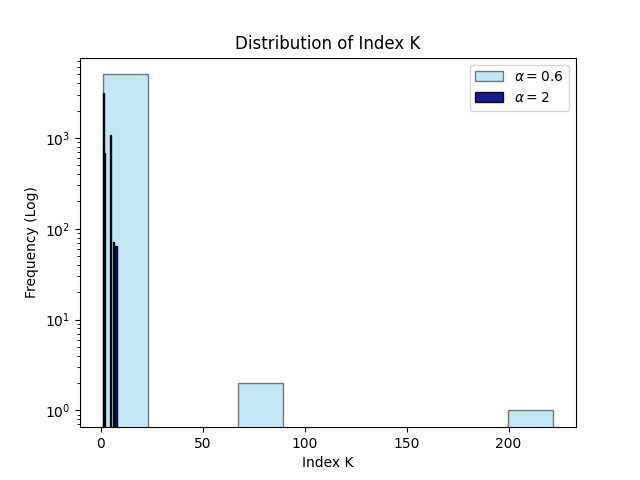}
    \includegraphics[width=0.47\linewidth]{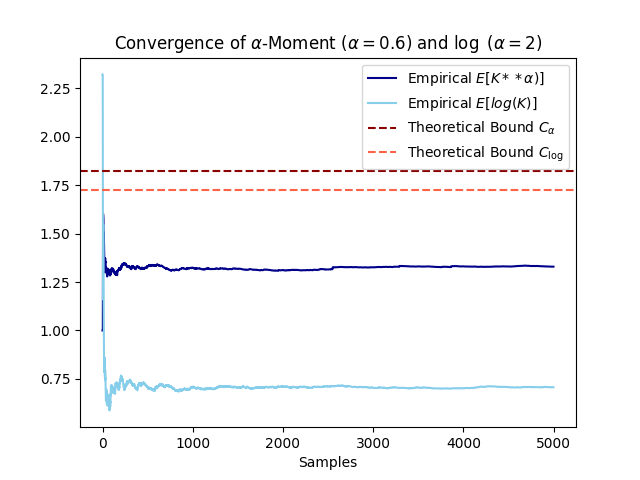}
    \caption{Numerical Validation of Rényi-SFRL ($\alpha = 0.6$ and $\alpha = 2$). Left: Histogram of the selected codebook indices $K$ (log scale), showing heavy-tailed behavior for $\alpha < 1 $ and finite support for $\alpha > 1$. Right: Convergence of the empirical $\alpha$-moment $\mathbb{E}[K^{0.6}]$ ($\mathbb{E}[\log (K)]$ for $\alpha = 2$, blue), remaining strictly below the theoretical bound (red).}
    \label{fig_sfrl_sim}
\end{figure}

We verify the moment constraint $\mathbb{E}[K^\alpha] \leq C_\alpha$ for $\alpha = 0.6$ and the logarithmic moment constraint $\mathbb{E}[ \log (K)] \leq C_{\log}$ for $\alpha = 2$, which together constitute the core of the achievability proof.
The theoretical bounds are calculated using Sibson's mutual information of order $\alpha+1$, with $C_\alpha = 2^{ \alpha I_{\alpha+1}(X;Y) } + \alpha$ for $\alpha = 0.6$ and the corresponding constant $C_{\log}$ for $\alpha=2$.
As shown in Fig.~\ref{fig_sfrl_sim}, the distribution of the selected indices $K$ exhibits the heavy-tailed behavior characteristic of the $\alpha < 1$ regime (dark blue), in contrast to the lighter-tailed distribution observed for $\alpha = 2$ (light blue). The cumulative empirical moment converges to approximately $1.329$ ($0.707$), which lies well below the theoretical upper bounds of $1.822$ ($1.726$) for $\alpha = 0.6$ ($\alpha = 2$, respectively). This confirms the validity of the Rényi-SFRL construction. 

\section{Conclusion}
\label{sec_conclude}
In this work, we generalize the RDP framework to the Rényi regime by adopting Sibson’s $\alpha$-mutual information as the fundamental complexity measure. This extension provides a principled way to analyze distortion–perception trade-offs beyond the Shannon setting and reveals new structural phenomena in both optimal representations and functional constructions.
Several directions for future work remain open. First, we aim to address the divergence in the regime $\alpha \in (0,0.5]$, where moment constraints alone do not ensure normalizability and may require additional regularization. Second, for $\alpha > 1$, we seek to tighten achievability bounds beyond the logarithmic moment constraint. Finally, we plan to apply these results to privacy-preserving and robust generative modeling under distribution shifts.


\bibliographystyle{IEEEtran}
\bibliography{ref}

\appendices
\section{Proof of Theorem \ref{thm_rd_gaussian}}
\label{apx:thm_rd_gaussian}
Sibson’s mutual information~\cite{Sibson1969} admits the following norm representation
\begin{align}
\label{eq_renyi_mi}
    I_\alpha(X; Y)=\frac{1}{\alpha-1}\log\|h\|_{1/\alpha},
\end{align}
where $h(y):=\int p_X(x)p_{Y|X}(y|x)^\alpha dx$.

\subsection*{Achievability (direct calculation with an affine Gaussian test channel)}
Fix $c \in \mathbb{R}$ and $Z \sim \mathcal{N}(0,\sigma_Z^2)$ be independent of $X$, and define $Y = cX + Z$.
The distortion is given by
\[
\mathbb{E}[(X-Y)^2] = (1-c)^2 \sigma_X^2 + \sigma_Z^2.
\]
Let $a := c^2\sigma_X^2$ and $b := \sigma_Z^2/\alpha$. Denote by $\phi_\sigma(x)$ the probability density function of a zero-mean Gaussian random variable with variance $\sigma_X^2$. Then
\begin{align}
    h(y)&=\int\phi_\sigma(x)\big(\phi_{\sigma_Z}(y-cx)\big)^\alpha dx \\
    &=(2\pi\sigma_Z^2)^{-\alpha/2}\frac{\sigma_Z}{\sqrt{\sigma_Z^2+\alpha a}}e^{-\frac{y^2}{2(a+b)}}. 
\end{align}
Consequently,
\[
\|h\|_{1/\alpha} =\Big(\int h(y)^{\frac{1}{\alpha}}dy\Big)^{\alpha}
=\alpha^{\frac{\alpha-1}{2}}\;\sigma_Z^{1-\alpha}\;(a+b)^{\frac{\alpha-1}{2}}.
\]
Substituting into~\eqref{eq_renyi_mi}, we obtain
\begin{align}
    &I_\alpha(X;Y)=\frac{1}{\alpha-1}\log\|h\|_{1/\alpha}\\
&=\frac12\log\!\Big(\alpha(a+b)\Big)-\frac12\log\sigma_Z^2
=\frac12\log\!\Big(1+\alpha\frac{c^2\sigma_X^2}{\sigma_Z^2}\Big).
\end{align}

We now minimize over $(c,\sigma_Z^2)$ subject to the distortion constraint $(1-c)^2\sigma_X^2 + \sigma_Z^2 = D$.
Equivalently, we maximize $r :=\frac{a}{\sigma_Z^2} = \frac{c^2\sigma_X^2}{D-(1-c)^2\sigma_X^2}$ over $c$ satisfying $(1-c)^2 \sigma_X^2 < D$. The maximizer is
\[
c^*=1-\frac{D}{\sigma_X^2},\quad {\sigma_Z^\star}^2=D\Big(1-\frac{D}{\sigma_X^2}\Big),\quad
r^*=\frac{\sigma_X^2-D}{D}.
\]
Substituting into $I_\alpha$ yields the achievable rate
\[
I_\alpha(X;Y) = \frac12\log\!\Big(1 + \alpha\frac{\sigma_X^2-D}{D}\Big).
\]

\subsection*{Converse}
\paragraph*{Step 1: Reduction to estimation followed by independent Gaussian noise}
For any channel $P_{Y|X}$ satisfying $\mathbb{E}[(X-Y)^2] \le D$, let $\tilde{X}:=\mathbb{E}[X\mid Y]$ denote the minimum mean-square estimator (MMSE) of $X$ given $Y$.
Then $\mathbb{E}[(X-\tilde{X})^2]\le D$, and by the DPI of Sibson's mutual information (valid for all $\alpha>0$),
\[
I_\alpha(X;\tilde{X})\le I_\alpha(X;Y).
\]
If $\Delta=\mathbb{E}[(X-\tilde{X})^2]<D$, introduce an independent Gaussian random variable $Z \sim \mathcal{N}(0,D-\Delta)$ and define $\tilde{Y}=\tilde{X}+Z$.
By the DPI, we have $I_\alpha(X;\tilde{Y})\le I_\alpha(X;\tilde{X})$, while the distortion constraint is met with equality: $\mathbb{E}[(X-\tilde{Y})^2] = D$.
Therefore, it suffices to lower bound $I_\alpha(X;\tilde{Y})$ over channels of the form
\[
\tilde{Y}=\tilde{X}+Z,\ \tilde{X}:=\mathbb{E}[X| Y],\ Z\sim\mathcal N(0,\sigma_Z^2),\ Z\perp (\tilde{X},X),
\]
where $Z$ is independent of $(\tilde{X}, X)$ and $\sigma_Z^2 = D - \Delta$.

\paragraph*{Step 2: Sharp Young with Gaussian extremality and optimization over exponents}
Let $f$ denote the probability density function of $\tilde{X}$ and let $k$ denote the probability density function of $Z$.
Then, 
\[
h(y)=\int f(u)k(y-u)^\alpha du=(f* k^\alpha)(y).
\]
Set $s = \frac{1}{\alpha}$ and for $p, q, s \ge 1$ satisfying $\frac1p + \frac1q = 1 + \frac{1}{s}$, the sharp Young inequality~\cite{Barthe1997sharpyoung} yields
\begin{align}
\label{eq_sharp_young_rhs}
    \|h\|_{s}\ \le\ C(p,q,s)\;\|f\|_{p}\;\|k^\alpha\|_{q},
\end{align}
where the Beckner--Brascamp--Lieb sharp constant is given by $C(p,q,s)=\frac{A_pA_q}{A_s}$,
$A_t=\big(t^{1/t}(t')^{-1/t'}\big)^{1/2}$, $t'=\frac{t}{t-1}$.
Define $u = 1/p'$ and $v = 1/q'$, so that $u + v = 1-\alpha$.
Equality in~\eqref{eq_sharp_young_rhs} holds if and only if both $f$ and $k^\alpha$ are Gaussian densities.
Using the Gaussian $L^r$-norm formula $\|\phi_{\sqrt v}\|_r=(2\pi)^{-(r-1)/(2r)} r^{-1/(2r)} v^{-1/(2r')}$, and
collecting constants, the right-hand side (RHS) of~\eqref{eq_sharp_young_rhs} can be expressed as
\begin{align}
    &C(p,q,s)\;\|f\|_{p}\;\|k^\alpha\|_{q}=\; K_\alpha (2\pi)^{-\frac{1-\alpha}{2}} s^{-\frac{1}{2s}} (s')^{\frac{1}{2s'}}\;
    \Psi(u,v),\\
    &\text{with}\ \Psi(u,v):=\frac{{\sigma_{\tilde{X}}^2}^{-\frac{u}{2}}\ \frac{\sigma_Z^2}{\alpha}^{-\frac{v}{2}}}{(u)^{-\frac{u}{2}}(v)^{-\frac{v}{2}}},
\end{align}
where $K_\alpha = (2\pi)^{\frac{1-\alpha}{2}}\alpha^{-\frac{1}{2}} \sigma_Z^{1-\alpha}$ is the scalar factor arising from $||k^\alpha\|_q$.

Denote $a = \sigma_{\tilde{X}}^2$ and $b = \frac{\sigma_Z^2}{\alpha}$. Optimizing $\Psi(u,v)$ under the constraint $u + v = 1-\alpha$ via the method of Lagrange multipliers yields
\[
u^*=\frac{(1-\alpha)a}{a+b},\ 
v^*=\frac{(1-\alpha)b}{a + b} \ \text{and} \  \Psi^\star = \left( \frac{1-\alpha}{a+b}\right)^{\frac{1-\alpha}{2}}.
\]
Hence, with $s = \frac{1}{\alpha}$and substituting the remaining constants in~\eqref{eq_sharp_young_rhs}, we obtain 
\begin{align}
\label{eq_rhs_sharpyoung}
    C(p,q,s)\;\|f\|_{p}\;\|k^\alpha\|_{q} &= K_\alpha (2\pi)^{-\frac{1-\alpha}{2}} s^{-\frac{1}{2s}} (s')^{\frac{1}{2s'}}\Psi^\star \notag\\
    &= \left(1 + \frac{\alpha\sigma_{\tilde{X}}^2}{\sigma_Z^2}\right)^{\frac{\alpha - 1}{2}}.
\end{align}

\paragraph*{Step 3: Lower bound on $I_{\alpha}$}
By the DPI for $I_\alpha$ and the bound in~\eqref{eq_rhs_sharpyoung}, we obtain 
\begin{align}
    I_\alpha (X; Y) \geq I_\alpha(X; \tilde{Y}) &= \frac{1}{\alpha - 1}\log{\left( ||f*g||_{\frac{1}{\alpha}}\right)} \\
    &= \frac{1}{2}\log\left(1 + \frac{\alpha\sigma_{\tilde{X}}^2}{\sigma_Z^2}\right),
\end{align}
where $g$ denotes the probability density function of $Z$.
This completes the proof of the converse.

\paragraph*{Step 4: Case $\alpha>1$} When $\alpha > 1$, we have $\alpha - 1 > 0$, and therefore minimizing \eqref{eq_renyi_mi} is equivalent to minimizing $\|h\|_{1/\alpha}$. By the reverse sharp Young inequality for exponents $0 < p, q, s < 1$, we obtain  
\begin{align}
    \|h\|_{s}\ \geq  C(p,q,s)\;\|f\|_{p}\;\|k^\alpha\|_{q}.
\end{align}
The remainder of the argument follows the same steps as in Step~2, leading to the desired lower bound.

\paragraph*{Step 5: Exact expression for Gaussian input}
Equality in all the preceding steps holds if and only if the following conditions are satisfied:
(i) $Y=\tilde{X}+Z$, where $Z$ is Gaussian and independent of $X$,  
(ii) $\tilde{X}$ is Gaussian with variance $c^2\sigma_X^2$. For a Gaussian source $X$, this implies that $(X, \tilde{X})$ is jointly Gaussian, and hence $\tilde{X} = cX$ for some scalar $c$;
(iii) under the MSE constraint $(1-c)^2\sigma_X^2 + \sigma_Z^2 = D$, the preceding analysis yields the optimal affine Gaussian test channel
\[
Y=cX + Z,\ c = 1 - \frac{D}{\sigma_X^2}, \sigma_Z^2 = D\left(1-\frac{D}{\sigma_X^2}\right).
\]
Consequently, the minimum Sibson coding rate is given by
\[
R_\alpha(D)=\frac12\log\!\Big(1+\alpha \frac{\sigma_X^2-D}{D}\Big),
\]
which completes the proof.

\section{Proof of Theorem \ref{thm_rdp_gaussian}}
\label{apx:thm_rdp_gaussian}
We now turn to the rate–distortion–perception problem defined in~\eqref{eq_rrdp} by incorporating the perception constraint, expressed through a divergence measure $d_P(P_X, P_Y)$.

\subsection{Parameterization}
From the previous section, we know that the linear additive Gaussian test channel is optimal under Sibson’s $\alpha$-mutual information. Under this test channel, the MSE distortion constraint becomes
\begin{align}
D
&= \mathbb E\big[(X-Y)^2\big]
= (1-c)^2\sigma_X^2 + \sigma_Z^2
= \sigma_X^2 + \sigma_{Y}^2 - 2c\sigma_X^2.
\end{align}
Hence, the following two parameters must be optimized under the distortion and perception constraints:
\begin{equation}
c=\frac{\sigma_X^2+\sigma_{Y}^2-D}{2\sigma_X^2},
\qquad
\sigma_Z^2=\sigma_{Y}^2-c^2\sigma_X^2.
\end{equation}
To ensure that the channel is valid (i.e., $\sigma_Z^2 \ge 0$), the reproduction variance $\sigma_Y^2$ must satisfy $(\sigma_X - \sigma_Y)^2 \le D$. We assume throughout that the problem constraints $D$ and $P$ admit a non-empty feasible set.

For $\alpha \neq 1$, the Rényi mutual information of the Gaussian channel is given by
\begin{equation}
I_\alpha(X;Y)
=\frac12\log\!\Big(1+\alpha \frac{c^2\sigma_X^2}{\sigma_Z^2}\Big)
=\frac12\log\!\Big(1+\alpha \mathrm{SNR}(\sigma_{Y}^2)\Big),
\end{equation}
where
\begin{equation}
\;
\mathrm{SNR}(\sigma_{Y}^2)
=\frac{(\sigma_X^2+\sigma_{Y}^2-D)^2}
{ 4\sigma_X^2\sigma_{Y}^2-(\sigma_X^2+\sigma_{Y}^2-D)^2 }. \;
\end{equation}
\begin{remark}
$\mathrm{SNR}(\sigma_{Y}^2)$ is minimized at $\sigma_X^2-D$, and is strictly
decreasing on the interval $[0,\sigma_X^2-D]$ and strictly increasing on $[\sigma_X^2-D,\infty)$.
\end{remark}

\subsection{Perception Constraint $\;d_P(P_{Y},P_X)\le P\;$ as an Interval in $\sigma_{Y}$}
When both marginals are zero-mean Gaussian, many perception measures $d_P$, such as the KL divergence, Wasserstein distance, and Rényi divergence, reduce to a one-dimensional constraint on the variance ratio $t:=\sigma_{Y}^2/\sigma_X^2$. This yields a feasible interval
\[
t \in [t_{\min}(P), t_{\max}(P)] \quad\Longleftrightarrow\quad
\sigma_{Y}\in\big[\underline{\sigma}_{Y}(P), \overline{\sigma}_{Y}(P)\big],
\]
where $\underline{\sigma}_{Y}=\sigma_X\sqrt{t_{\min}}$ and $\overline{\sigma}_{Y}=\sigma_X\sqrt{t_{\max}}$.

Since $I_\alpha(X; Y)$ is monotone in $\mathrm{SNR}(\sigma_{Y}^2)$ and the latter attains its minimum at $\sigma_X^2-D$, the optimal reproduction variance is given by the projection of $\sigma_X^2-D$ onto the feasible set, namely 
\begin{equation}
{\sigma_{Y}^\star}^2 = \min \left\{ \max\left\{\sigma_X^2 - D,  \underline{\sigma}_{Y}^2\right\}, \  \overline{\sigma}_{Y}^2\right\}.
\end{equation}
The corresponding channel parameters are
\begin{equation}
c^\star=\frac{\sigma_X^2+{\sigma_{Y}^\star}^2-D}{2\sigma_X^2},\qquad
{\sigma_{Z}^\star}^2={\sigma_{Y}^\star}^2-{c^\star}^2\sigma_X^2,
\end{equation}
and the minimum rate for distortion level $D$ and perception constraint $P$ is given by
\begin{equation}
\begin{aligned}
    R_\alpha(D,P)&=\frac12\log \left(1+\alpha \frac{{c^\star}^2\sigma_X^2}{{\sigma_{Z}^\star}^2}\right)
    \\
    &=\frac12\log \left(1+\alpha \mathrm{SNR}({\sigma_{Y}^\star}^2)\right).
\
\end{aligned}
\end{equation}

\subsection{Examples}
We now provide examples of perception measures $d_P$.

\subsubsection*{2-Wasserstein distance $W_2$}
For one-dimensional zero-mean Gaussian distributions, we have
\[
W_2\big(\mathcal{N}(0,\sigma_{Y}^2),\mathcal{N}(0,\sigma_X^2)\big)=|\sigma_{Y}-\sigma_X|\le P.
\]
Consequently, the feasible interval for the reproduction standard deviation is given by
\[
{\;
\underline{\sigma}_{Y}(P)=\max\{0, \sigma_X-P\},\qquad \overline{\sigma}_{Y}(P)=\sigma_X+P.
\;}
\]

\subsubsection*{KL divergence $D_{\mathrm{KL}}$}
For zero-mean Gaussian distributions, the KL divergence is given by
\[
D_{\mathrm{KL}}\big(\mathcal{N}(0,\sigma_{Y}^2) \| \mathcal{N}(0,\sigma_X^2)\big)
=\frac12\Big(t-1-\log t\Big)\ \le\ P.
\]
The function $t\mapsto t-1-\log t$ is strictly decreasing on $(0,1]$ and strictly increasing on $[1,\infty)$, and attains its minimum value $0$ at $t = 1$. Hence, there exists a unique solution $t_{\min}(P) \in (0,1)$ to $\tfrac12(t_{\min}-1-\log t_{\min}) = P$, and a corresponding solution $t_{\max}(P)\in [1, \infty)$.
Therefore, the feasible interval for the reproduction standard deviation is
\[
{\;
\underline{\sigma}_{Y}(P)=\sigma_X \sqrt{t_{\min}(P)},\quad
\overline{\sigma}_{Y}(P)={\sigma_X}\sqrt{t_{\max}(P)}.
\;}
\]

\subsubsection*{Rényi divergence $D_\beta$ (order $\beta>0,\ \beta\neq1$)}
For zero-mean Gaussian distributions, the Rényi divergence of order $\beta$ is given by
\[
D_\beta\big(\mathcal{N}(0,\sigma_{Y}^2)\big\|\mathcal{N}(0,\sigma_X^2)\big)
=\frac{1}{2(\beta-1)}
\log\left(\frac{(1-\beta)+\beta t}{t^\beta}\right).
\]
For each $\beta$, the mapping $t \mapsto D_\beta(t)$ is strictly decreasing on $(0,1]$ and strictly increasing on $[1,\infty)$, with $D_\beta(1) = 0$. Hence, the feasible set is given by
\[
t\in\big[t_{\min}^{(\beta)}(P),t_{\max}^{(\beta)}(P)\big],
\qquad t_{\max}^{(\beta)}(P)=\frac{1}{t_{\min}^{(\beta)}(P)}.
\]
Therefore, the feasible interval for the reproduction standard deviation is
\[
{\;
\underline{\sigma}_{Y}(P)=\sigma_X\sqrt{t_{\min}^{(\beta)}(P)},\qquad
\overline{\sigma}_{Y}(P)=\frac{\sigma_X}{\sqrt{t_{\min}^{(\beta)}(P)}}.
\;}
\]

\section{Proof of Theorem~\ref{thm_renyi_sfrl}}
\label{apx_sfrl}
In this appendix, we provide the detailed derivation of the moment constraint $\mathbb{E}[K^\alpha] \leq C_\alpha$ for the case $\alpha \in (0, 1)$, as stated in Theorem~\ref{thm_renyi_sfrl}. This bound is central to our analysis, as it establishes that the heavy-tailed coding cost is controlled by Sibson's mutual information of order $\alpha+1$. We also provide the corresponding derivation for the case $\alpha > 1$, where the result relies on a logarithmic moment bound.

\subsection*{Construction via Poisson functional representation}
We employ the Poisson Functional Representation using the minimizing distribution $Q_Y$. Let $Z = \{(\tilde{Y}_i, T_i)\}_{i \geq 1}$ be a Poisson point process on $\mathcal{Y} \times \mathbb{R}_{\geq 0}$ with intensity measure $Q_Y \times \nu$, where $\nu$ denotes the Lebesgue measure. 
Note that we do not take $Q_Y = P_Y$; instead, we choose $Q_Y = Q_{Y_\alpha}$, which is optimal for characterizing $I_\alpha(X; Y)$.
Given a source realization $X = x$, the functional representation index $K$ is selected as
\begin{equation}
K = \argmin_{i \geq 1} \frac{T_i}{L(Y_i|x)}, \ \text{where } L(y|x) = \frac{dP_{Y|X}(y|x)}{dQ_{Y_\alpha}(y)}.
\end{equation}
By the displacement theorem for Poisson point processes~\cite{kingman1992poisson}, setting
$Y = \tilde{Y}_K$ ensures that $Y \sim P_{Y|X}(\cdot|x)$. Thus, $Y$ is a measurable function of the source $X$ and the shared randomness $Z$, indexed by $K$. 

The proof relies on the generalized Poisson matching lemma~\cite{li_sfrl_tit}[Lemma~3], which bounds concave functions of the index $K$ in terms of the likelihood ratio $L$.
\begin{lemma}
[Poisson Matching Lemma for Concave Functions]
\label{lemma_pml}
Let $g: \mathbb{R}^+ \to \mathbb{R}^+$ be a non-decreasing, concave function. Then, conditioned on $X=x$, the expected cost of the index $K$ satisfies
\begin{equation}
\mathbb{E}[g(K) \mid X=x] \leq \mathbb{E}_{Y \sim P_{Y|X}(\cdot|x)} \left[ g\left( L(Y|x) + 1 \right) \right].
\end{equation}
\end{lemma}
By invoking the generalized Poisson matching lemma~\cite{li_sfrl_tit} for any concave, non-decreasing function $g$, we obtain 
\begin{equation}
\begin{aligned}
    &\mathbb{E}[g(K) \mid X=x] \leq \mathbb{E}_{Y \sim P_{Y|X}(\cdot|x)}\left[ g\left( \frac{dP_{Y|X}(Y|x)}{dQ_{Y_\alpha}(Y)} + 1 \right) \right] \\
    &\le \mathbb{E}_{Y \sim P_{Y|X}(\cdot|x)}\left[ g\left( \frac{dP_{Y|X}(Y|x)}{dQ_{Y_\alpha}(Y)}  \right) +g'\left( \frac{dP_{Y|X}(Y|x)}{dQ_{Y_\alpha}(Y)}\right)\right],
\end{aligned}
\end{equation}
where the second inequality follows from the concavity of $g$.

\subsection*{Case 1: $0<\alpha<1$}
For $\alpha\in (0, 1)$, the function $g(u) = u^\alpha$ is concave. Setting $g(u) = u^\alpha$ in Lemma~\ref{lemma_pml}, we obtain
\begin{equation}
\begin{aligned}
    &\mathbb{E}[K^\alpha \mid X=x] \\
    &\overset{(a)}{\le} \mathbb{E}_{Y \sim P_{Y|X}(\cdot|x)}\left[  L(Y|x)  ^\alpha + \alpha  L(Y|x)^{\alpha -1}\right]\\
    &\overset{(b)}{=}\mathbb{E}_{Y \sim P_{Y|X}(\cdot|x)}\left[ L(Y|x)^\alpha \right] + \alpha \mathbb{E}_{Y \sim Q}\left[ L(Y|x)^{\alpha }\right ]\\
    &\overset{(c)}{\leq} 2^{\alpha D_{\alpha+1}(P_{Y|X}(\cdot|x)||Q_{Y_{\alpha}})} + \alpha,
\end{aligned}
\end{equation}
where $(a)$ follows from Lemma~\ref{lemma_pml} applied to the concave function $g(u)=u^\alpha$, $(b)$ follows from a change of measure, and $(c)$ follows from Jensen’s inequality and the definition of the Rényi divergence of order $\alpha+1$.

Taking the expectation with respect to $P_X$~\cite{verdu2015alpha}, we obtain the global moment constraint $C_\alpha$:
\begin{equation}
    \mathbb{E}[K^\alpha] \leq 2^{\alpha I_{\alpha+1}(X; Y)} + \alpha \triangleq C_\alpha.
\end{equation}
This bound shows that, although the index $K$ may exhibit heavy-tailed behavior for $\alpha<1$, its $\alpha$-moment remains finite and is explicitly controlled by the Sibson mutual information $I_{\alpha+1}(X;Y)$.

This moment constraint is directly connected to the variational characterization of Sibson’s $\alpha$-mutual information, specifically through its interpretation in terms of collision probabilities. To upper bound the entropy, we consider the maximum Rényi entropy problem subject to an $\alpha$-moment constraint:
\begin{equation}
\label{eq_lagrange_1}
\begin{aligned}
    &\max_{p_k} \frac{1}{1-\alpha} \log \left( \sum_{k=1}^\infty p_k^\alpha \right) \\
    \text{subject to: }& \ \sum_{k=1}^\infty p_k k^\alpha \leq C_\alpha, \  \sum_k p_k=1.
\end{aligned}
\end{equation}
Since $\alpha \in (0, 1)$, maximizing $H_\alpha$ is equivalent to maximizing the concave functional $\sum p_k^\alpha$. This leads to the following Lagrangian, with multipliers $\lambda$ and $\mu$ enforcing the moment and normalization constraints:
\begin{align}
    &\quad \mathcal{L}(p_k, \lambda, \mu) \notag \\&= \sum_k p_k^{\alpha} - \lambda(\sum_k p_k k^{\alpha}) - C_\alpha) - \mu (\sum_k p_k - 1)
\end{align}
Taking derivatives with respect to $p_k$ and applying the KKT conditions yields
\begin{equation}
    p_k^{\alpha-1} = \frac{\lambda k^\alpha + \mu }{\alpha},
\end{equation}
which leads directly to~\eqref{eq_renyi_1}. 

As $k \to \infty$, the probability mass function exhibits the asymptotic decay $p_k \sim k^{\frac{\alpha}{\alpha-1}}$. 
For the distribution to be normalizable (i.e., $\sum_k p_k = 1$), the decay exponent must satisfy $\frac{\alpha}{1-\alpha} > 1$, which is equivalent to $\alpha > \tfrac{1}{2}$. Under this condition, the Lagrange multipliers $\lambda$ and $\mu$ can be chosen to satisfy both the moment constraint and the normalization constraint.

The optimal distribution maximizing Rényi entropy under the $\alpha$-moment constraint has infinite support with a polynomially decaying tail. This implies that representations in the $\alpha < 1$ regime are inherently heavy-tailed, requiring a conceptually infinite codebook to achieve the theoretical bound.

\subsection*{Case 2: $\alpha > 1$}
In this regime, the function $u^\alpha$ is convex, which precludes a direct upper bound via the Poisson matching lemma. Instead, we bound the logarithmic moment $\mathbb{E}[\log K]$, as in the classical Shannon case. It was shown in~\cite{li2024pointwiseredundancyoneshotlossy} that
\begin{equation}
    \mathbb{E}[\log K \mid X=x] \leq D(P_{Y|X}(\cdot|x) \| Q_Y) + 1.
\end{equation}
Averaging over $X$, we obtain the following bound on the expected log-length:
\begin{equation}
    \mathbb{E}[\log K] \leq I(X;Y)+D(P_Y||Q_Y) + 1 \triangleq C_{log}.
\end{equation}

We now solve the maximum Rényi entropy problem for $\alpha > 1$, given by
\begin{equation}
\begin{aligned}
    &\max_{p_k} \frac{1}{1-\alpha} \log \left( \sum_{k=1}^\infty p_k^\alpha \right) \\
    \text{subject to:} &\ \sum_{k=1}^\infty p_k \log k \leq C_{\log}, \ \sum_k p_k=1.
\end{aligned}
\end{equation}
For $\alpha > 1$, the prefactor $\frac{1}{1-\alpha}$ is negative, and therefore maximizing $H_\alpha$ is equivalent to minimizing the $\sum p_k^\alpha$. Solving the corresponding Lagrangian optimization problem yields
\begin{equation}
\label{eq_apx_pk_2}
    p_k^{\alpha-1} =\left[\frac{-(\lambda \log k + \mu)}{\alpha} \right]^+,
\end{equation}
This expression implies that for $\alpha > 1$, the optimal distribution has finite support, and hence the strong functional representation compresses the shared randomness into a finite alphabet, whose size is governed by Sibson’s mutual information. Moreover, using~\eqref{eq_apx_pk_2}, we directly recover the bound stated in~\eqref{eq_renyi_2}.

The optimal distribution under the log-moment constraint has strictly finite support. In particular, the probability $p_k$ vanishes for all $k > K_{\max}$ (since $\lambda \log k + \mu > 0$ in this regime), where $K_{\max} \approx 2^{I(X;Y)+ D_{KL}(P_Y||Q_Y)}$. This implies that for $\alpha > 1$, the strong functional representation effectively compresses the source into a finite alphabet, whose size is governed by the mutual information (up to a divergence term).

\end{document}